\documentclass[twocolumn]{revtex4-1}

\usepackage{amsmath}    % need for subequations
\usepackage{graphicx}   % need for figures
\usepackage{epsfig}
\usepackage{algorithmic}
\usepackage{bm}% bold math

\begin{document}

\title{Self Avoiding Paths Routing Algorithm in Scale-Free Networks}

\author{Abdeljalil Rachadi}
   \email{rachadi\_abdeljalil@yahoo.fr}
\author{ Mohamed Jedra}
 \email{jedra@fsr.ac.ma}
\author{Noureddine Zahid}
  \email{zahid@fsr.ac.ma}
\affiliation{ Laboratoire Conception et Syst\`emes (Mico\'electronique et Informatique),\\
Facult\'e des Sciences, Universit\'e Mohammed V, Agdal,\\ Avenue Ibn Batouta, B.P. 1014, Rabat 10000, Morocco.}

\date{May 21 2012}

\begin{abstract}
In this paper, we present a new routing algorithm called "the Self Avoiding Paths Routing Algorithm". Its application to traffic flow in sclae-free networks shows a great improvement over the so called "efficient routing" protocol while at the same time maintaining a relatively low average packet travel time. It has the advantage of minimizing path overlapping throughout the network in a self consistent manner with a relatively small number of iterations by maintaining an equilibrated path distribution especially among the hubs. This results in a significant shifting of the critical packet generation rate over which traffic congestion occurs, thus permitting the network to sustain more information packets in the free flow state. The performance of the algorithm is discussed both on a Bar\'abasi-Albert (BA) network and real autonomous system (AS) network data.
\end{abstract}
\maketitle

\begin{quotation}
The understanding of information flow throughout a complex network is an important issue from both a theoretical and a practical point of view. The goal of such understanding is to control the traffic in complex systems such as the internet. Indeed, the main problem with these systems is the emergence of congestion when the traffic load becomes higher than a certain threshold. To solve this problem, studies focus on the mechanism of routing of information packets to reach their destinations. In this respect, the well known shortest path protocol directs the traffic more likely towards the more connected nodes (hubs). Many alternative routing protocols (static and traffic aware protocols) have been proposed in order to obtain an equilibrated traffic load between the different paths. The latter protocols need extra communication between the routers compared to the static ones which can easily be implemented. But the proposed static algorithms, did not, in general, use the optimal paths explicitly. By introducing a self avoiding mechanism between the paths during the process of their construction, we were able to devise a new routing protocol called: "the Self Avoiding Paths Routing Algorithm" which results in an optimal distribution of the paths and performs better than many of the previous protocols in raising the overall network capacity.
\end{quotation}

\section{INTRODUCTION}

The routing problem represents one of the central questions in today's internet traffic engineering \cite{Ramasco,Zhaolai,Lawniczak,Bamatraf,Salles}. Different routing protocols lead to different outcomes concerning the traffic flow throughout the network. In this respect, we can distinguish three classes of routing schemes \cite{YinZhou,YanZhou,LingHu,LingHu_pherom}. The first class uses a fixed routing strategy where the paths depend only on the physical structure of the network and are stored in a fixed routing table of each router. The second class consists of dynamic routing protocols where the decision to send a packet through a given link depends only on the traffic state. Finally, in the third class, both the physical structure and the dynamic traffic information are used in order to take routing decisions \cite{echenique,demartino,holme}. First methods present many advantages in terms of economical and technical costs compared with traffic aware methods and are widely used in mid and small systems \cite{tanenbaum}.

The traffic flow in communication networks is characterized by a phase transition from the free flow state, where the packet generation rate is balanced by the packet delivery rate, to the congested state where packets accumulate rapidly in the network. This happens when the traffic generation rate $R$ becomes greater than a threshold $R_{c}$. The principal task of traffic engineering is to make $R_{c}$ as high as possible in order that the network could sustain much traffic load without congestion.

In this respect, the shortest path routing (SPR) used for forwarding information packets\cite{ericsson,fortz} is not sufficient for this goal. Indeed, the structure of complex networks \cite{newman} such as the internet \cite{PasVesp,PPJ} is best described by an underlying scale-free \cite{BA,caldarelli,zhaopark,herrero,ZhaoCupertino,HuangChow} complex structure characterized by the existence of hubs (highly connected nodes) as well as peripheral nodes. Routing based on shortest paths often tends to send more information packets towards the hubs which rapidly become congested resulting in general congestion of the whole network.

Other routing schemes have been proposed to overcome this difficulty. The so called efficient path routing (EPR) (ref.~\onlinecite{YanZhou}) was used instead of SPR, with the cost of a path is chosen to depend on the degrees of its nodes: $\sum_{i}k_i^{\beta}$ (where $k_i$ is the degree of node i and $\beta$ is a control parameter). The best results are obtained with this algorithm for $\beta=1$, though at the expense of the average packet travel time. On the other hand, in the hub avoidance (HA) routing\cite{sreenivasan}, the authors remove, in a first step, some hubs from the network which becomes a set of disconnected clusters, and then build usual shortest paths between every pair of nodes in the same cluster before placing the hubs back and computing the paths between the remaining pairs of nodes. The results are found to be better than those obtained when using the shortest path (SPR) protocol. 

Although theses algorithms succeed in raising the critical packet generation rate, they do not explicitly take into account the true optimal paths that should be used for routing. Another algorithm called "Optimal routing" (OR)\cite{danilla2006,danilla2007} was proposed which tries to reduce the maximum node betweenness in a self-consistent manner reaching a near-optimal configuration of paths. Good results are obtained by this algorithm despite its slow convergence.

In this paper, we propose a new algorithm which minimizes self-consistently, the number of intersection nodes between the resulting paths. The results of this "Self Avoiding Paths Routing" (SAPR) algorithm show improvements in comparison to the EPR presented in Ref.~\onlinecite{YanZhou} while at the same time reaching near-optimal path configuration in a relatively small number of iterations compared to the OR (Refs.~\onlinecite{danilla2006,danilla2007}). In fact, the paths generated by our algorithm are not "self avoiding" in the strict sens; the SAPR algorithm tries to find self avoiding paths in an optimal sens, that is, paths that are as much as possible "self avoiding" or with minimal intersection points.

\section{MODEL}
The principal idea behind the SAPR algorithm is to execute iteratively the standard Dijkstra's algorithm for finding shortest paths between any given pair of source and destination nodes with the following main modification: instead of using the cost associated with each path as a fixed input for a Dijkstra's cycle, this cost is updated even during a cycle whenever a new path is discovered. This adaptive process permits us to take into account the previous situation in order to decide how to constract subsequent paths, by trying to avoid the existing ones for the remaining path finding process during the same cycle. The output costs of one iteration are used as an input for the following one. 

A judicious choice for the cost associated with a path ${\cal P}_{ij}$ from node $i$ to node $j$ can be chosen according to the following form:
\begin{equation}
W({{\cal P}_{ij}})=\sum_{u\in{\cal P}_{ij}}w(u)
\end{equation} 
where $w(u)$ represents the cost associated with node $u$ given by
\begin{equation}
\label{cost}
w(u)={[{\cal N}_{p}(u)}]^\alpha
\end{equation} 
where ${\cal N}_{p}(u)$ is the number of paths found by the algorithm in the current step and traversing node $u$. $\alpha$ is an adjustable parameter. For $\alpha=0$, $W({{\cal P}_{ij}})$ is just the number of nodes in the path ${\cal P}_{ij}$, and so we recover the shortest path algorithm.
${\cal N}_{p}(u)$ is also calles the node betweenness (see ref.~\onlinecite{sreenivasan}).

The SAPR algorithm evolves just like the Dijkstra's one. We will not give the details of this algorithm, we will only show how to compute the costs associated with each node during the path construction process. In the following, we will call a "shortest path" a path that will have the lowest cost.

Network nodes are visited in the same order used in the standard Dijkstra's algorithm. In order to explore all shortest paths starting from a given source node $s$ to all the remaining $N-1$ nodes, the algorithm starts at node $s$ and records the "distance" $d(v)$ from $s$ to a any node $v$. This distance is given by the total cost of the nodes which constitute the path ${\cal P}_{sv}$:
\begin{equation}
d(v)=\sum_{u\in{\cal P}_{sv}}w(u)
\end{equation}
Whenever a path with the lowest cost is discovered from $s$ to a node $v$ through an immediately preceeding node $u$, we need to store $u$ as the antecedent of $v$: ${\cal A}_s(v)=u$. For the sake of simplicity, we will not cover the case where more than one shortest path is stored in the routing table, but will consider a unique path between any given pair of nodes. The generalization is straightforward.

\begin{figure}[th]
\centerline{\psfig{file=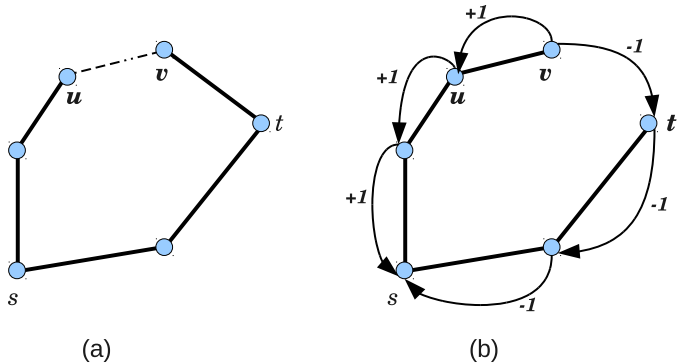,width=3.5in}}
\vspace*{8pt}
\caption{\label{pathconst}Path discovery process of the SAPR algorithm, (a): The algorithm is in node $u$ and tries to check for node $v$ with an already existing path via $t$. (b): In this case, the path through $u$ is found to have the lowest cost and the path through $t$ is removed, so the node costs are updated for the two paths down to the source node $s$. The operations $+1$ and $-1$ are to be performed on the number of paths on every node along the two paths (see the two pseudo-codes in text).}
\end{figure}

Suppose that the algorithm is now on the current node $u$, and that a neighbor $v$ of $u$ is checked (see Fig.~\ref{pathconst}). The node $v$ may have already a previously discovered lower cost path from source $s$: ${\cal P}^t_{sv}$, through another node $t$ with a corresponding distance: $d_t(v)=\sum_{j\in{\cal P}^t_{sv}}w(j)$. So we are faced with the three following situations:

\textit{Case 1}: if $d_t(v)<d(u)+w(v)$, then the path through $t$ still has the lowest cost, so nothing is done concerning node $v$ in this step.

\textit{Case 2}: if $d_t(v)>d(u)+w(v)$, then the the distance to $v$ from $u$ is discovered as the new lower cost path. In this case, we need to update the number of paths passing through all nodes of the path ${\cal P}^u_{sv}$ and the corresponding costs by traversing the path backward from node $v$ down to source node $s$ and adding $1$ to the number of paths of every node belonging to the path as is shown by the following pseudo-code:

\algsetup{indent=2em}
\begin{algorithmic}
\STATE $r\gets u$
\WHILE {$r\ne s$} 
   \STATE ${\cal N}_{p}(r)\gets {\cal N}_{p}(r)+1$
   \STATE $w(r)\gets{[{\cal N}_{p}(r)}]^\alpha$
   \STATE $r\gets A_{s}(r)$
\ENDWHILE
\end{algorithmic}

In the other hand, all nodes of the previous path ${\cal P}^t_{sv}$ should be updated according to:

\begin{algorithmic}
\STATE $r\gets t$
\WHILE {$r\ne s$} 
   \STATE ${\cal N}_{p}(r)\gets {\cal N}_{p}(r)-1$
   \STATE $w(r)\gets{[{\cal N}_{p}(r)}]^\alpha$
   \STATE $r\gets A_{s}(r)$
\ENDWHILE
\end{algorithmic}
meaning that a path traversing the corresponding nodes is no longer used. So, before the updating, the path from node $s$ to node $v$ via node $t$ has the lowest cost: ${\cal P}_{sv}={\cal P}^t_{sv}$. After being updated, the path through node $u$ is discovered as the new lower cost one: ${\cal P}_{sv}={\cal P}^u_{sv}$.

\textit{Case 3}: if $d_t(v)=d(u)+w(v)$, then the two paths through $u$ and through $t$ are of equal cost, so we choose either case~1 or case~2 randomly (with probability $1/2$) in the case where we want to store in the routing table a unique shortest path between any pair of nodes. But in the contrary, the path through ${\cal P}^t_{sv}$ has to be stored as a shortest path as well.

The algorithm proceeds according to the following steps:
\begin{enumerate}
\item{Assign 1 to the cost of every node.}
\item{Compute the shortest paths from a given source node to every destination node; and for every path found, update the number of paths for the corresponding nodes and compute the costs following the two pseudo-codes above. }
\item{Repeat step 2 without initializing the number of paths nor the costs associated with every node but with using them as input fot the next cycle.}
\end{enumerate}

\section{SIMULATION RESULTS AND DISCUSSION}

\begin{figure}[th]
\centerline{\psfig{file=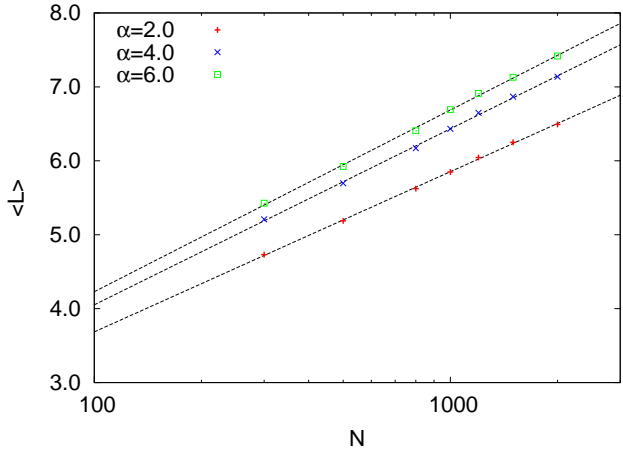,width=3.5in}}
\vspace*{8pt}
\caption{\label{small_world}Plot of the average path length $\langle L \rangle$ as a function of the number $N$ of nodes in the BA network for $\alpha=2.0, 4.0$ and $6.0$.}
\end{figure}

\begin{figure}[th]
\centerline{\psfig{file=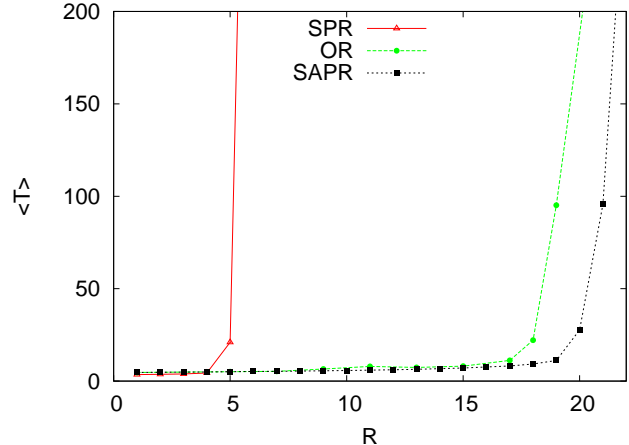,width=3.5in}}
\vspace*{8pt}
\caption{\label{comp_OR}Plot of the average travel time $\langle T \rangle$ as a function of the packet generation rate $R$ for the algorithms SPR and OR compared with the SAPR algorithm with $\alpha=6.0$  for the BA network with $N=200$ nodes.}
\end{figure}

For most of our numerical simulations, we used a Bar\'abasi-Albert (BA) network (Ref.~\onlinecite{BA}) with $N=1000$ nodes (except for fig.~\ref{small_world} where we used values up to $N=2000$ and fig.~\ref{comp_OR} where $N=200$). This network is built starting with $m_0=3$ fully connected nodes; and at each time step, a new node with $m=2$ edges is added to the existing nodes with preferential attachment, that is with probability $p_i$ that depends on the node degree $k_i$ of every candidate node:
\begin{equation}
p_i=\frac{k_i}{\sum_j{k_j}}
\end{equation}
The network generated has an average node degree: $\langle k \rangle=2m=4$. Furthermore, it has a power-law degree distribution characterized by the existence of highly connected nodes (hubs) and peripheral nodes.

In addition, in order to test the proposed algorithm on different types of complex networks, we have done traffic simulations on a real autonomous system (AS) network data which consist of periodic snapshots of BGP routing table dumps~\footnote{see website: http://www.routeviews.org}. For our purpose, we used a network of size $N=549$ from the AS-733 dataset.

\begin{figure}[th]
\centerline{\psfig{file=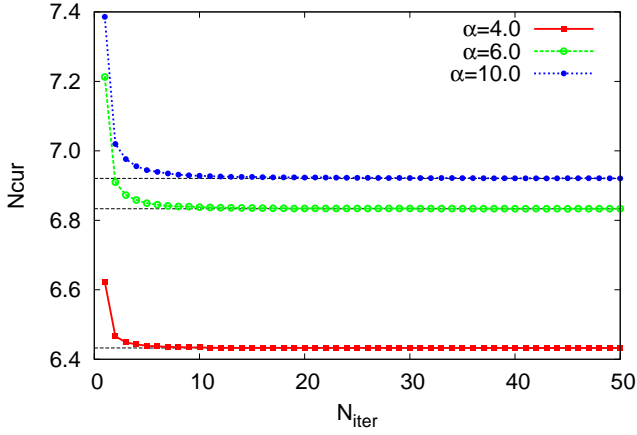,width=3.5in}}
\vspace*{8pt}
\caption{\label{path_iter}Number of current packets in the network $N_{cur}$ as a function of the number of iterations $N_{iter}$ for the SAPR algorithm for different values of the parameter $\alpha$ (BA network with $N=1000$).}
\end{figure}

Using the BA network with the SAPR algorithm, we first check for the number of necessary iterations to obtain the convergence of the algorithm. The average path length is calculated and its variation is reported in Fig.~\ref{path_iter} in function of the number of iterations for $\alpha=4.0, 6.0$ and $10.0$ for the BA network. It can be seen that the algorithm quickly converges to a near-optimal value within just a few iterations. The simulations show that for small values of $\alpha$, the convergence is very fast; for example, when $0<\alpha \leq 3$, a number of $15$ iterations suffices while for $3<\alpha \leq 10$, we need up to $50$ iterations. However, when $\alpha$ becomes larger (strong interacting paths), we need much more iterations, but in return, there is no substantial gain in terms of the performance as we will show later.

In Fig.~\ref{small_world}, we have plotted the average path length $\langle L \rangle$ as a function of the number of nodes in the BA network: $N$ for $\alpha=2.0, 4.0$ and $6.0$. We can see that, even if $\langle L \rangle$ is a monotonically increasing function of $N$,  the small world property\cite{kleinberg} is maintained for different values of $\alpha$, that is:
\begin{equation}
\label{small_world_eq}
\langle L \rangle \sim \log N.
\end{equation}

We used the SAPR algorithm to simulate traffic flow in the BA network where each node is treated as a host and a router, as follows: For each time step, we generate $R$ packets from random sources and assign them random destinations, then put each of them at the end of a queue on the corresponding source. At each time step, every node can deliver at most $C$ packets from the top of its queue towards their destinations (Here we set $C=1$ without loss of generality). The packets are navigated one step forward to their destinations by using the fixed routing table maintained at each router. If a packet's destination is found among the neighbors of the current node, it is directly sent to it; else, it is sent to the neighboring node with the minimal distance to the destination of the packet and is put at the bottom of its queue. Packets reaching their destinations are automatically removed from the system.

\begin{figure}[th]
\centerline{\psfig{file=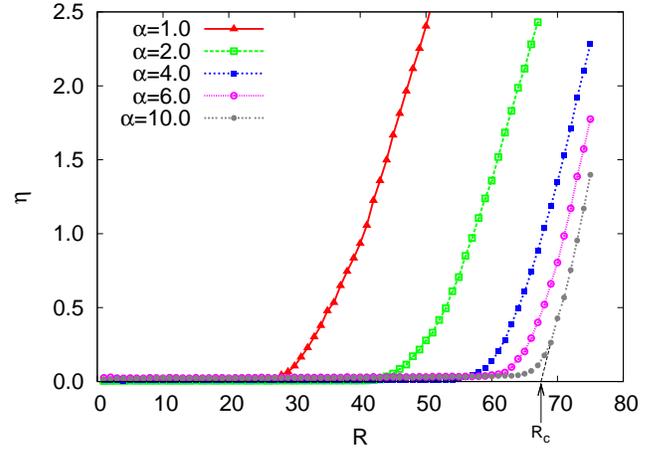,width=3.5in}}
\vspace*{8pt}
\caption{\label{ord_param}Order parameter $\eta$ (given by Eq.~\ref{order_parameter}) of the SAPR algorithm as a function of the packet generation rate $R$ for different $\alpha$ values. The arrow indicates the position of the critical value $R_c \approx 67$ for $\alpha=10.0$ (BA network with $N=1000$).}
\end{figure}

To characterize the traffic dynamics, we used the order parameter $\eta$ given by\cite{arenas}:
\begin{equation}
\label{order_parameter}
\eta=\lim_{t\to \infty}\frac{C}{R}\frac{\langle N_{p}(t+\Delta t)-N_p(t) \rangle}{\Delta t}
\end{equation}

where $N_p(t)$ is the number of packets in the network at time $t$ and $\langle ... \rangle$ is the average over time windows of width $\Delta t$. In Fig.~\ref{ord_param}, we report the order parameter $\eta$ versus $R$, the number of packets generated per unit time for $\alpha=1.0, 2.0, 4.0, 6.0$ and $10.0$. We can distinguish two phases, for $R<R_c$, where $R_c$ is a critical value, the system is in a free flow state where the number of generated packets is balanced by the number of packets delivered to their destinations and $\eta=0$. While for $R>R_c$, a congested phase takes place on the network and $\langle N_{p}(t+\Delta t)-N_p(t) \rangle$ grows linearly with $\Delta t$ and thus $\eta$ tends towards a constant value for fixed $C$ and $R$ values. It is clear that $R_c$ increases with $\alpha$ and we can conclude that the SAPR algorithm is very efficient for larger values of $\alpha$. This result is consistent with the fact that when $\alpha$ is large, the paths tend to be as much separated as possible. So when the information packets use these paths, they, more likely, try to generate an equally distributed load among the different routers (regardless if they are hubs or peripheral nodes) resulting in more traffic load in the free flow state.

\begin{figure}[th]
\centerline{\psfig{file=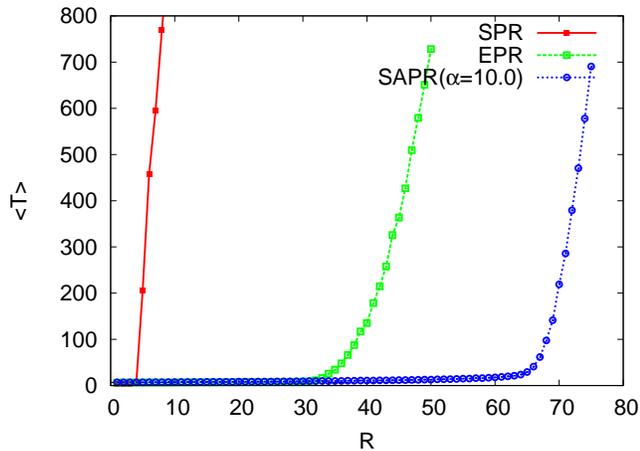,width=3.5in}}
\vspace*{8pt}
\caption{\label{compar}Plot of the average travel time $\langle T \rangle$ as a function of the packet generation rate $R$ for the algorithms SPR and EPR compared with the SAPR algorithm with $\alpha=10.0$ (BA network with $N=1000$).}
\end{figure}

\begin{figure}[th]
\centerline{\psfig{file=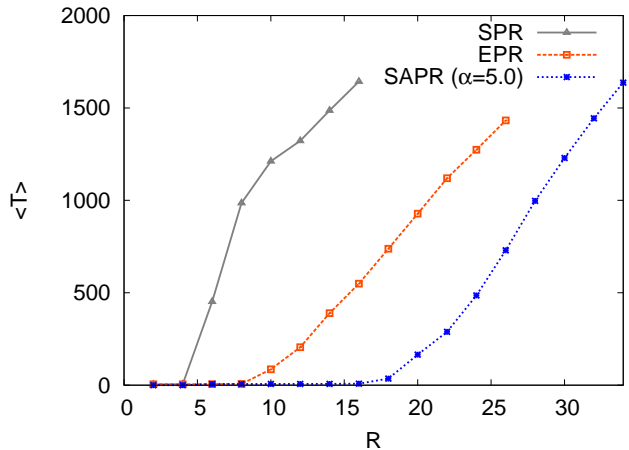,width=3.5in}}
\vspace*{8pt}
\caption{\label{compar_real}Plot of the average travel time $\langle T \rangle$ as a function of the packet generation rate $R$ for the algorithms SPR and EPR compared with the SAPR algorithm with $\alpha=5.0$, for the AS-733 Oregon Route Views dataset network with $N=549$.}
\end{figure}

In order to compare the performance of the SAPR algorithm (for $\alpha=10.0$) with the shortest path routing (SPR) and the efficient path routing (EPR)(Ref.~\onlinecite{YanZhou}) algorithms, we have plotted in Fig.\ref{compar}, the variation of the average packet travel time $\langle T \rangle$ in function of $R$ for the three algorithms which was found to have the same behaviour as the order parametet $\eta$ and the same transition point at $R_c$. We can see clearly the large value of $R_c$ for the SAPR algorithm compared with the others which is an indication that it is much more performant than SPR and EPR protocols.

The average travel time is also plotted in Fig.\ref{compar_real} for the algorithm: SPR, EPR and SAPR (with $\alpha=5.0$) for the AS-733 network. The results clearly confirms the fact that the SAPR algorithm shows its superiority not only for the theoretical BA networks, but also on real network data.

\begin{figure}[th]
\centerline{\psfig{file=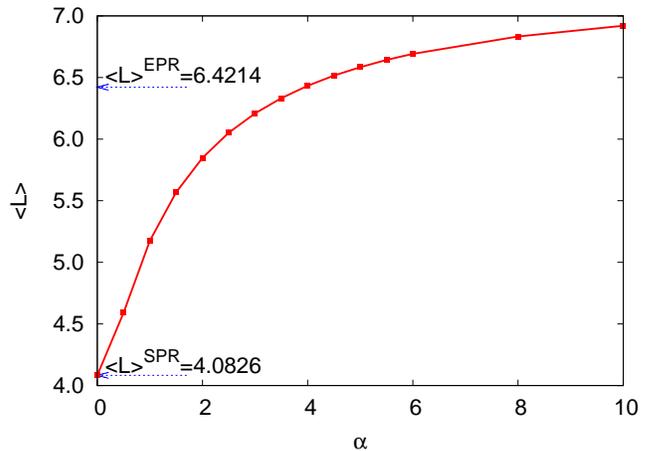,width=3.5in}}
\caption{\label{avlength}Variation of the average path length $\langle L \rangle$ vs. $\alpha$. The values of $\langle L \rangle$ corresponding to the algorithms SPR and EPR are indicated for comparison by the arrows (BA network with $N=1000$).}
\end{figure}

\begin{figure}[th]
\centerline{\psfig{file=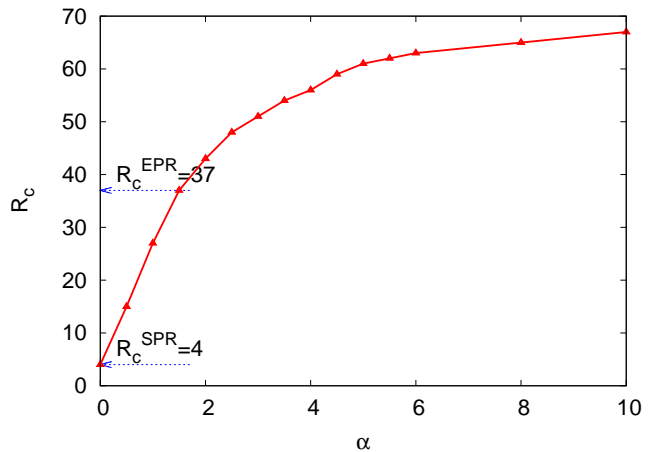,width=3.5in}}
\vspace*{8pt}
\caption{\label{gen_rate}Critical packet generation rate $R_c$ as a function of $\alpha$ values. The arrows indicate the values corresponding to the SPR and the EPR algorithms (BA network with $N=1000$).}
\end{figure}

Moreover, we found that the average travel time $\langle T \rangle$ is relatively lower for the SAPR algorithm even in the free flow phase for some range of $\alpha$ values. It is straightforward to see that the average travel time is equal to the average path length $\langle L \rangle$ for low values of $R$: $\langle L \rangle=\lim_{R\to 0}\langle T \rangle$ in the case where the sources and destinations of packets are both chosen at random. In Fig.~\ref{avlength}, we have reported $\langle L \rangle$ as a function of $\alpha$. We can see that $\langle L \rangle$ increases to a limiting value corresponding to $\alpha\to\infty$. This is due to the fact that when $\alpha=0$, the SAPR protocol is equivalent to the SPR one and the average path length has its absolute minimal value because of the abscence of any "interaction" between the paths that are free of any constraint during their construction process. When $\alpha$ increases, however, the mutual avoidance between the paths becomes more important and they are more and more constrained to look for less occupied nodes and thus become longer and try to choose their optimal configuration where the average path length reaches its maximum.

\begin{figure*}[th]
\centerline{\psfig{file=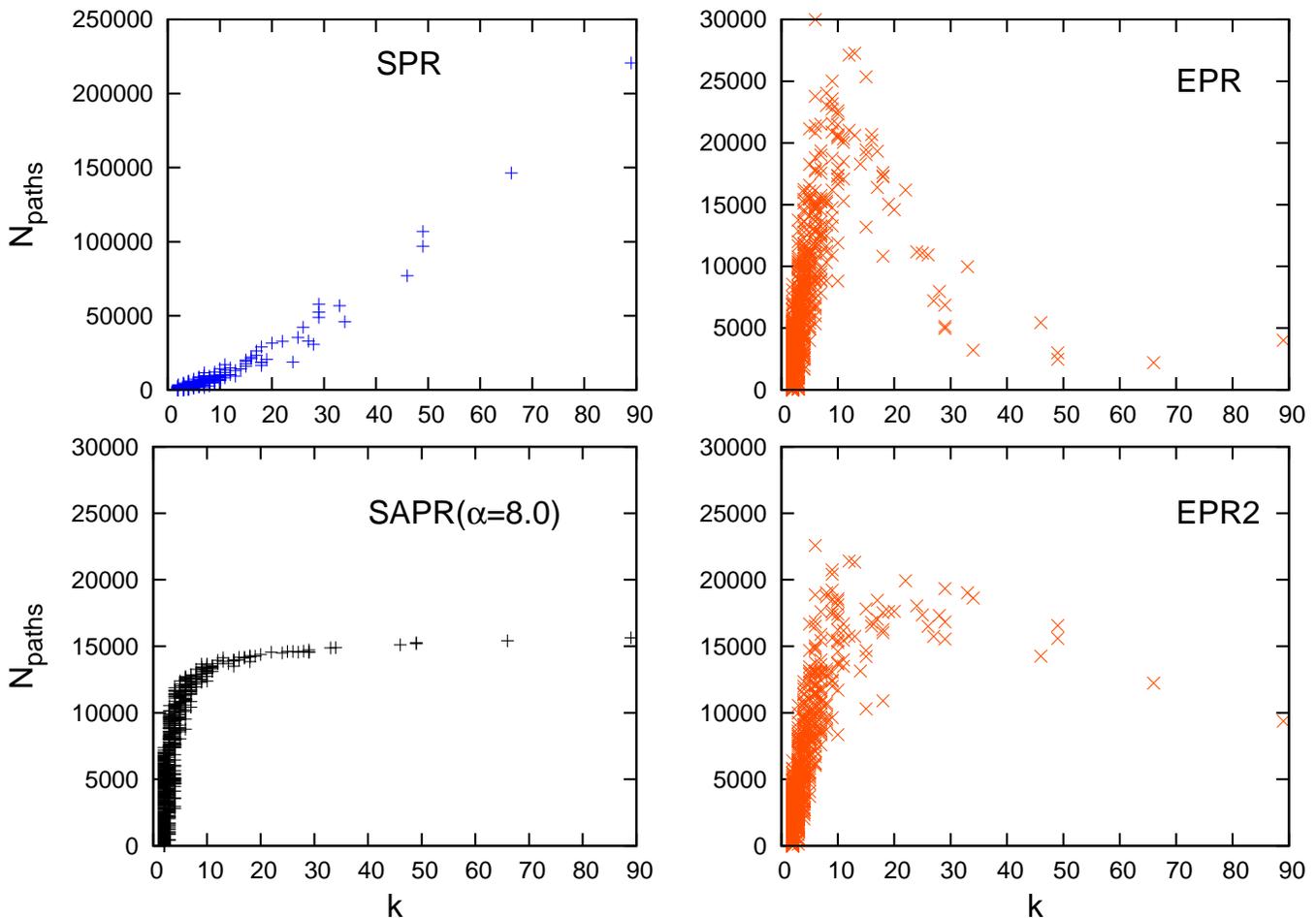,width=7.0in}}
\vspace*{8pt}
\caption{\label{nb_paths}Dependence of the number of paths $N_{paths}$ traversing each node on the node degree $k$ for the three algorithms: SPR, EPR, SAPR (with $\alpha=8.0$) and EPR2 (BA network with $N=1000$).}
\end{figure*}

Although the average path length becomes somewhat longer, this is compensated by a larger value of the critical packet generation rate $R_c$ (see Fig.~\ref{gen_rate}) which has the same behaviour as $\langle L \rangle$ with a low starting value for $\alpha=0$ corresponding to the SPR algorithm: $R_c^{SPR}\approx4$ to the value $R_c^{SAPR}\approx67$ for $\alpha=10$ which is larger even than the EPR algorithm: $R_c^{EPR}\approx37$. For further greater values of $\alpha$, the increase of $R_c$ is very weak, and the gain in performance for the traffic flow is not important compared to the large number of iterations necessary for convergence. 

Now we will try to explain the performance of the SAPR algorithm compared especially to the EPR one. To this end, we plotted in Fig.~\ref{nb_paths}, the number of paths $N_{paths}$ traversing each node (or its betweenness) as a function of the node degree $k$. As can be expected, the SPR protocol results in a large number of paths traversing the hubs compared to the peripheral nodes. In contrast, the EPR protocol succeeds in reducing the number of paths traversing the hubs while nodes with intermediate degree receive more paths. But we believe that this algorithm reduces the number of paths crossing the hubs more than necessary. Indeed, as the figure shows, the SAPR algorithm establishes a more equilibrated distribution of the paths among the hubs.

To further clarify our argument, we have modified the cost in the EPR algorithm ($w(i)=k_i$ see Ref.~\onlinecite{YanZhou}) by reducing the cost of the hubs (having $k_i>k_c$) with $k_c=15$ for example, at the expense of the other nodes, by using the simple formula $w(i)=k_i$ for $k_i<k_c$ and $w(i)=k_c^{\epsilon}.k_i^{1-\epsilon}$ otherwise ($\epsilon$ is a parameter in the range $[0,1]$ and the factor $k_c^{\epsilon}$ guarantees continuity). For $\epsilon=0.3$, this algorithm (called EPR2) performs better than the standard EPR reaching $R_c\approx 43$ instead of $37$ for the same BA network. This result was achieved by the fact that the paths distribution is more equilibrated than that of the EPR protocol (see Fig.~\ref{nb_paths}), though not optimal like the SAPR protocol.

In another respect, although the SAPR algorithm presents some similarities with the OR one, in the fact that they are both heuristic algorithms that try iteratively to find a near optimal solution for the routing problem by acting on the node betweenness, we will discuss here some differences between them. While the SAPR algorithm tries to redistribute paths across the whole network by acting on the node betweenness of all the network nodes in every move during every iteration of the algorithm, the OR protocol, in the other hand, tries to reshape the betweenness landscape by minimizing the maximum betweenness ${\cal B}_{max}$ by updating it at the end of each iteration.

From a computational point of view, the OR protocol runs in $\mathcal{O} (N^3 \log N)$ (see Refs.\onlinecite{danilla2007})). In the other hand, the SAPR is based on the Dijkstra's algorithm which runs in $\mathcal{O} (M+N \log N)$ (if implemented with the Fibonacci heap), in order to find all the shortest paths starting from a given source node, where $N$ and $M$ are, respectively, the number of nodes and the number of links in the network. The SAPR protocol uses the supplementing step of traversing the paths backwards to update their costs, so the time becomes $\mathcal{O} (M \langle L \rangle +N \log N)$ ($\langle L \rangle$ is the average path length). Taking into account the fact that the paths are constructed for the $N$ sites and that the simulation takes $\mathcal{O} (N)$ iterations, we conclude that the SAPR algorithm takes $\mathcal{O} (N^2(M \langle L \rangle+N \log N))$ execution time.

For networks that obey the small world property (Eq.~\ref{small_world_eq}), such as the BA network for example and the fact that $M$ is proportional to $N$ in this type of networks, we are left with the final running time of the SAPR algorithm: $\mathcal{O} (N^3 \log N)$ in the worst case. Then, we can say that SAPR and the OR algorithms (see ref.~\onlinecite{danilla2007}) perform in relatively the same computational time in the large networks limit.

For a direct quantitative comparision of the two algorithms, we have plotted in Fig.~\ref{comp_OR}, the variation of the average travel time for the SPR, the OR and the SAPR protocols for the BA network with $N=200$ nodes. The results for the critical packet generation rates are respectively $R_c^{SPR}\sim5$, $R_c^{OR}\sim18$ and $R_c^{SAPR}\sim20$, showing some advantage of the SAPR protocol.

In the other hand, although the two algorithms have the same execution time in the large network limit, we have experienced a very slow convergence using the OR protocol wich needs up to $2000$ iterations to obtain a satisfactory convergence for the maximum betweenness (for the BA network with $200$ nodes). While to obtain convergence of the SAPR protocol, we need only $\sim15$ iterations (see for example Fig.~\ref{path_iter}). One possible explanation to this difference is the fact that the OR is a form of extremal optimization which tries to reduce the quantity ${\cal B}_{max}$ on a single node in every step, whereas the SAPR takes advantage of the path interaction on each node during every single iteration and uses this collective information for optimization, thus reaching quickly an equilibrated path distribution landscape accross the network.

We note here that we have done extensive traffic simulation for the goal of explicit comparison between the performance of the different algorithms in real traffic situation. But if the mere goal is to determine the transition point from free flow to congested state, it can be achieved directly from the maximal betweenness ${\cal B}_{max}$ through the relation (see ref~.\onlinecite{sreenivasan}): $\gamma_c=(N-1)/{\cal B}_{max}$, where $\gamma_c$ is the critical probability of packet insertion rate at a given node.

\section{CONCLUSION}
In summary, we have proposed a new routing algorithm called the Self Avoiding Path Routing (SAPR) algorithm where the paths are self-avoiding in the sens that the inetersection between them is reduced at its minimum. It uses the number of paths crossing each node of a given path as the cost associated with that path. Theses costs are found self-consistently by updating them during each step of the path discovery process of the original Dijksta's algorithm. This fact results in a more equilibrated path distribution especially among the hubs.

The direct result was the increase of the critical packet generation rate above which a jamming state occurs resulting in a much greater traffic load that can be sustained by the network without beeing congested. Moreover, the average path length was found to be relatively small and obeys the small world property.

The algorithm performs better than both the shortest path routing "SPR" and the efficient path routing "EPR" protocols. The former was shown to have the majority of paths concentrated on the hubs while the latter was shown to avoid hubs more than necessary. It was shown also that the SAPR algorithm takes the same computational time as the OR protocol but this latter have a slower convergence.
\nocite{*}
%\bibliography{article}

\begin{thebibliography}{30}%
\makeatletter
\providecommand \@ifxundefined [1]{%
 \@ifx{#1\undefined}
}%
\providecommand \@ifnum [1]{%
 \ifnum #1\expandafter \@firstoftwo
 \else \expandafter \@secondoftwo
 \fi
}%
\providecommand \@ifx [1]{%
 \ifx #1\expandafter \@firstoftwo
 \else \expandafter \@secondoftwo
 \fi
}%
\providecommand \natexlab [1]{#1}%
\providecommand \enquote  [1]{``#1''}%
\providecommand \bibnamefont  [1]{#1}%
\providecommand \bibfnamefont [1]{#1}%
\providecommand \citenamefont [1]{#1}%
\providecommand \href@noop [0]{\@secondoftwo}%
\providecommand \href [0]{\begingroup \@sanitize@url \@href}%
\providecommand \@href[1]{\@@startlink{#1}\@@href}%
\providecommand \@@href[1]{\endgroup#1\@@endlink}%
\providecommand \@sanitize@url [0]{\catcode `\\12\catcode `\$12\catcode
  `\&12\catcode `\#12\catcode `\^12\catcode `\_12\catcode `\%12\relax}%
\providecommand \@@startlink[1]{}%
\providecommand \@@endlink[0]{}%
\providecommand \url  [0]{\begingroup\@sanitize@url \@url }%
\providecommand \@url [1]{\endgroup\@href {#1}{\urlprefix }}%
\providecommand \urlprefix  [0]{URL }%
\providecommand \Eprint [0]{\href }%
\providecommand \doibase [0]{http://dx.doi.org/}%
\providecommand \selectlanguage [0]{\@gobble}%
\providecommand \bibinfo  [0]{\@secondoftwo}%
\providecommand \bibfield  [0]{\@secondoftwo}%
\providecommand \translation [1]{[#1]}%
\providecommand \BibitemOpen [0]{}%
\providecommand \bibitemStop [0]{}%
\providecommand \bibitemNoStop [0]{.\EOS\space}%
\providecommand \EOS [0]{\spacefactor3000\relax}%
\providecommand \BibitemShut  [1]{\csname bibitem#1\endcsname}%
\let\auto@bib@innerbib\@empty
%</preamble>
\bibitem [{\citenamefont {Ramasco}\ \emph {et~al.}(2005)\citenamefont
  {Ramasco}, \citenamefont {de~la Lama}, \citenamefont {L\'opez},\ and\
  \citenamefont {Boettcher}}]{Ramasco}%
  \BibitemOpen
  \bibfield  {author} {\bibinfo {author} {\bibfnamefont {J.~J.}\ \bibnamefont
  {Ramasco}}, \bibinfo {author} {\bibfnamefont {M.~S.}\ \bibnamefont {de~la
  Lama}}, \bibinfo {author} {\bibfnamefont {E.}~\bibnamefont {L\'opez}}, \ and\
  \bibinfo {author} {\bibfnamefont {S.}~\bibnamefont {Boettcher}},\ }\href@noop
  {} {\bibfield  {journal} {\bibinfo  {journal} {{Phys.\ Rev. E}}\ }\textbf
  {\bibinfo {volume} {2010}},\ \bibinfo {pages} {036119} (\bibinfo {year}
  {2005})}\BibitemShut {NoStop}%
\bibitem [{\citenamefont {Zhao}\ \emph {et~al.}(2005)\citenamefont {Zhao},
  \citenamefont {Lai}, \citenamefont {Park},\ and\ \citenamefont
  {Ye}}]{Zhaolai}%
  \BibitemOpen
  \bibfield  {author} {\bibinfo {author} {\bibfnamefont {L.}~\bibnamefont
  {Zhao}}, \bibinfo {author} {\bibfnamefont {Y.~C.}\ \bibnamefont {Lai}},
  \bibinfo {author} {\bibfnamefont {K.}~\bibnamefont {Park}}, \ and\ \bibinfo
  {author} {\bibfnamefont {N.}~\bibnamefont {Ye}},\ }\href@noop {} {\bibfield
  {journal} {\bibinfo  {journal} {{Phys.\ Rev. E}}\ }\textbf {\bibinfo {volume}
  {71}},\ \bibinfo {pages} {026125} (\bibinfo {year} {2005})}\BibitemShut
  {NoStop}%
\bibitem [{\citenamefont {Lawniczak}\ and\ \citenamefont
  {Tang}(2006)}]{Lawniczak}%
  \BibitemOpen
  \bibfield  {author} {\bibinfo {author} {\bibfnamefont {A.~T.}\ \bibnamefont
  {Lawniczak}}\ and\ \bibinfo {author} {\bibfnamefont {X.}~\bibnamefont
  {Tang}},\ }\href@noop {} {\bibfield  {journal} {\bibinfo  {journal} {{Eur.
  Phys.\ J. B}}\ }\textbf {\bibinfo {volume} {50}},\ \bibinfo {pages} {231}
  (\bibinfo {year} {2006})}\BibitemShut {NoStop}%
\bibitem [{\citenamefont {Bamatraf}\ and\ \citenamefont
  {Othmana}(2007)}]{Bamatraf}%
  \BibitemOpen
  \bibfield  {author} {\bibinfo {author} {\bibfnamefont {M.}~\bibnamefont
  {Bamatraf}}\ and\ \bibinfo {author} {\bibfnamefont {M.}~\bibnamefont
  {Othmana}},\ }\href@noop {} {\bibfield  {journal} {\bibinfo  {journal}
  {Computer Communications}\ }\textbf {\bibinfo {volume} {30}},\ \bibinfo
  {pages} {1513} (\bibinfo {year} {2007})}\BibitemShut {NoStop}%
\bibitem [{\citenamefont {Salles}\ and\ \citenamefont {Rolla}(2007)}]{Salles}%
  \BibitemOpen
  \bibfield  {author} {\bibinfo {author} {\bibfnamefont {R.~M.}\ \bibnamefont
  {Salles}}\ and\ \bibinfo {author} {\bibfnamefont {V.~G.}\ \bibnamefont
  {Rolla}},\ }\href@noop {} {\bibfield  {journal} {\bibinfo  {journal}
  {Computer Communications}\ }\textbf {\bibinfo {volume} {30}},\ \bibinfo
  {pages} {1942} (\bibinfo {year} {2007})}\BibitemShut {NoStop}%
\bibitem [{\citenamefont {Yin}\ \emph {et~al.}(2006)\citenamefont {Yin},
  \citenamefont {Wang}, \citenamefont {Wang}, \citenamefont {Zhou},\ and\
  \citenamefont {Yang}}]{YinZhou}%
  \BibitemOpen
  \bibfield  {author} {\bibinfo {author} {\bibfnamefont {C.-Y.}\ \bibnamefont
  {Yin}}, \bibinfo {author} {\bibfnamefont {B.-H.}\ \bibnamefont {Wang}},
  \bibinfo {author} {\bibfnamefont {W.-X.}\ \bibnamefont {Wang}}, \bibinfo
  {author} {\bibfnamefont {T.}~\bibnamefont {Zhou}}, \ and\ \bibinfo {author}
  {\bibfnamefont {H.-J.}\ \bibnamefont {Yang}},\ }\href@noop {} {\bibfield
  {journal} {\bibinfo  {journal} {Physics Letters A}\ }\textbf {\bibinfo
  {volume} {351}},\ \bibinfo {pages} {220} (\bibinfo {year}
  {2006})}\BibitemShut {NoStop}%
\bibitem [{\citenamefont {Yan}\ \emph {et~al.}(2006)\citenamefont {Yan},
  \citenamefont {Zhou}, \citenamefont {Hu}, \citenamefont {Fu}, ,\ and\
  \citenamefont {Wang}}]{YanZhou}%
  \BibitemOpen
  \bibfield  {author} {\bibinfo {author} {\bibfnamefont {G.}~\bibnamefont
  {Yan}}, \bibinfo {author} {\bibfnamefont {T.}~\bibnamefont {Zhou}}, \bibinfo
  {author} {\bibfnamefont {B.}~\bibnamefont {Hu}}, \bibinfo {author}
  {\bibfnamefont {Z.~Q.}\ \bibnamefont {Fu}}, , \ and\ \bibinfo {author}
  {\bibfnamefont {B.~H.}\ \bibnamefont {Wang}},\ }\href@noop {} {\bibfield
  {journal} {\bibinfo  {journal} {Phys.\ Rev. E}\ }\textbf {\bibinfo {volume}
  {73}},\ \bibinfo {pages} {046108} (\bibinfo {year} {2006})}\BibitemShut
  {NoStop}%
\bibitem [{\citenamefont {Ling}\ \emph {et~al.}(2010)\citenamefont {Ling},
  \citenamefont {Hu}, \citenamefont {Jiang},\ and\ \citenamefont
  {Wu}}]{LingHu}%
  \BibitemOpen
  \bibfield  {author} {\bibinfo {author} {\bibfnamefont {X.}~\bibnamefont
  {Ling}}, \bibinfo {author} {\bibfnamefont {M.-B.}\ \bibnamefont {Hu}},
  \bibinfo {author} {\bibfnamefont {R.}~\bibnamefont {Jiang}}, \ and\ \bibinfo
  {author} {\bibfnamefont {Q.-S.}\ \bibnamefont {Wu}},\ }\href@noop {}
  {\bibfield  {journal} {\bibinfo  {journal} {Phys.\ Rev. E}\ }\textbf
  {\bibinfo {volume} {81}},\ \bibinfo {pages} {016113} (\bibinfo {year}
  {2010})}\BibitemShut {NoStop}%
\bibitem [{\citenamefont {Ling}\ \emph {et~al.}(2009)\citenamefont {Ling},
  \citenamefont {Hu}, \citenamefont {Jiang}, \citenamefont {Wang},
  \citenamefont {Cao},\ and\ \citenamefont {Wu}}]{LingHu_pherom}%
  \BibitemOpen
  \bibfield  {author} {\bibinfo {author} {\bibfnamefont {X.}~\bibnamefont
  {Ling}}, \bibinfo {author} {\bibfnamefont {M.-B.}\ \bibnamefont {Hu}},
  \bibinfo {author} {\bibfnamefont {R.}~\bibnamefont {Jiang}}, \bibinfo
  {author} {\bibfnamefont {R.}~\bibnamefont {Wang}}, \bibinfo {author}
  {\bibfnamefont {X.-.}\ \bibnamefont {Cao}}, \ and\ \bibinfo {author}
  {\bibfnamefont {Q.-S.}\ \bibnamefont {Wu}},\ }\href@noop {} {\bibfield
  {journal} {\bibinfo  {journal} {Phys.\ Rev. E}\ }\textbf {\bibinfo {volume}
  {80}},\ \bibinfo {pages} {066110} (\bibinfo {year} {2009})}\BibitemShut
  {NoStop}%
\bibitem [{\citenamefont {Echenique}, \citenamefont {G{\'o}mez-Garde{\~n}es},\
  and\ \citenamefont {Moreno}(2004)}]{echenique}%
  \BibitemOpen
  \bibfield  {author} {\bibinfo {author} {\bibfnamefont {P.}~\bibnamefont
  {Echenique}}, \bibinfo {author} {\bibfnamefont {J.}~\bibnamefont
  {G{\'o}mez-Garde{\~n}es}}, \ and\ \bibinfo {author} {\bibfnamefont
  {Y.}~\bibnamefont {Moreno}},\ }\href@noop {} {\bibfield  {journal} {\bibinfo
  {journal} {Physical Review E}\ }\textbf {\bibinfo {volume} {70}},\ \bibinfo
  {pages} {056105} (\bibinfo {year} {2004})}\BibitemShut {NoStop}%
\bibitem [{\citenamefont {Martino}\ \emph {et~al.}(2009)\citenamefont
  {Martino}, \citenamefont {Dall\'Asta}, \citenamefont {Bianconi},\ and\
  \citenamefont {Marsili}}]{demartino}%
  \BibitemOpen
  \bibfield  {author} {\bibinfo {author} {\bibfnamefont {D.~D.}\ \bibnamefont
  {Martino}}, \bibinfo {author} {\bibfnamefont {L.}~\bibnamefont {Dall\'Asta}},
  \bibinfo {author} {\bibfnamefont {G.}~\bibnamefont {Bianconi}}, \ and\
  \bibinfo {author} {\bibfnamefont {M.}~\bibnamefont {Marsili}},\ }\href@noop
  {} {\bibfield  {journal} {\bibinfo  {journal} {Phys.\ Rev. E}\ }\textbf
  {\bibinfo {volume} {79}},\ \bibinfo {pages} {015101} (\bibinfo {year}
  {2009})}\BibitemShut {NoStop}%
\bibitem [{\citenamefont {Holme}(2003)}]{holme}%
  \BibitemOpen
  \bibfield  {author} {\bibinfo {author} {\bibfnamefont {P.}~\bibnamefont
  {Holme}},\ }\href {doi:10.1142/S0219525903000803} {\bibfield  {journal}
  {\bibinfo  {journal} {Advances In Complex Systems}\ }\textbf {\bibinfo
  {volume} {6}},\ \bibinfo {pages} {163} (\bibinfo {year} {2003})}\BibitemShut
  {NoStop}%
\bibitem [{\citenamefont {Tanenbaum}(1996)}]{tanenbaum}%
  \BibitemOpen
  \bibfield  {author} {\bibinfo {author} {\bibfnamefont {A.~S.}\ \bibnamefont
  {Tanenbaum}},\ }\href@noop {} {\emph {\bibinfo {title} {{Computer
  Networks}}}}\ (\bibinfo  {publisher} {{Prentice Hall Press}},\ \bibinfo
  {year} {1996})\BibitemShut {NoStop}%
\bibitem [{\citenamefont {Ericsson}, \citenamefont {Resende},\ and\
  \citenamefont {Pardalos}(2002)}]{ericsson}%
  \BibitemOpen
  \bibfield  {author} {\bibinfo {author} {\bibfnamefont {M.}~\bibnamefont
  {Ericsson}}, \bibinfo {author} {\bibfnamefont {M.~G.~C.}\ \bibnamefont
  {Resende}}, \ and\ \bibinfo {author} {\bibfnamefont {P.~M.}\ \bibnamefont
  {Pardalos}},\ }\href@noop {} {\bibfield  {journal} {\bibinfo  {journal}
  {Journal of Combinatorial Optimization}\ }\textbf {\bibinfo {volume} {6}},\
  \bibinfo {pages} {299} (\bibinfo {year} {2002})}\BibitemShut {NoStop}%
\bibitem [{\citenamefont {Fortz}\ and\ \citenamefont {Thorup}(2002)}]{fortz}%
  \BibitemOpen
  \bibfield  {author} {\bibinfo {author} {\bibfnamefont {B.}~\bibnamefont
  {Fortz}}\ and\ \bibinfo {author} {\bibfnamefont {M.}~\bibnamefont {Thorup}},\
  }\href@noop {} {\bibfield  {journal} {\bibinfo  {journal} {IEEE Journal on
  Selected Areas in Communications}\ }\textbf {\bibinfo {volume} {20}},\
  \bibinfo {pages} {756} (\bibinfo {year} {2002})}\BibitemShut {NoStop}%
\bibitem [{\citenamefont {{M. E. J. Newman}}(2003)}]{newman}%
  \BibitemOpen
  \bibfield  {author} {\bibinfo {author} {\bibnamefont {{M. E. J. Newman}}},\
  }\href@noop {} {\bibfield  {journal} {\bibinfo  {journal} {SIAM Review}\
  }\textbf {\bibinfo {volume} {45}},\ \bibinfo {pages} {167} (\bibinfo {year}
  {2003})}\BibitemShut {NoStop}%
\bibitem [{\citenamefont {Pastor-Satorras}\ and\ \citenamefont
  {Vespignani}(2004)}]{PasVesp}%
  \BibitemOpen
  \bibfield  {author} {\bibinfo {author} {\bibfnamefont {R.}~\bibnamefont
  {Pastor-Satorras}}\ and\ \bibinfo {author} {\bibfnamefont {A.}~\bibnamefont
  {Vespignani}},\ }\href@noop {} {\emph {\bibinfo {title} {{Evolution and
  Structure of the Internet: A Statistical Physics Approach}}}}\ (\bibinfo
  {publisher} {{Cambridge University Press}},\ \bibinfo {address} {Cambridge},\
  \bibinfo {year} {2004})\BibitemShut {NoStop}%
\bibitem [{\citenamefont {Paul}, \citenamefont {Pan},\ and\ \citenamefont
  {Jain}(2011)}]{PPJ}%
  \BibitemOpen
  \bibfield  {author} {\bibinfo {author} {\bibfnamefont {S.}~\bibnamefont
  {Paul}}, \bibinfo {author} {\bibfnamefont {J.}~\bibnamefont {Pan}}, \ and\
  \bibinfo {author} {\bibfnamefont {R.}~\bibnamefont {Jain}},\ }\href@noop {}
  {\bibfield  {journal} {\bibinfo  {journal} {Computer Communications}\
  }\textbf {\bibinfo {volume} {34}},\ \bibinfo {pages} {2} (\bibinfo {year}
  {2011})}\BibitemShut {NoStop}%
\bibitem [{\citenamefont {{A.-L. Barab\'asi and R. Albert}}(1999)}]{BA}%
  \BibitemOpen
  \bibfield  {author} {\bibinfo {author} {\bibnamefont {{A.-L. Barab\'asi and
  R. Albert}}},\ }\href@noop {} {\bibfield  {journal} {\bibinfo  {journal}
  {Science}\ }\textbf {\bibinfo {volume} {286}},\ \bibinfo {pages} {509}
  (\bibinfo {year} {1999})}\BibitemShut {NoStop}%
\bibitem [{\citenamefont {{G. Caldarelli}}(2007)}]{caldarelli}%
  \BibitemOpen
  \bibfield  {author} {\bibinfo {author} {\bibnamefont {{G. Caldarelli}}},\
  }\href@noop {} {\emph {\bibinfo {title} {{Scale-free Networks: Complex Webs
  in Nature and Technology}}}}\ (\bibinfo  {publisher} {{Oxford University
  Press}},\ \bibinfo {address} {Oxford},\ \bibinfo {year} {2007})\BibitemShut
  {NoStop}%
\bibitem [{\citenamefont {Zhao}, \citenamefont {Park},\ and\ \citenamefont
  {Lai}(2004)}]{zhaopark}%
  \BibitemOpen
  \bibfield  {author} {\bibinfo {author} {\bibfnamefont {L.}~\bibnamefont
  {Zhao}}, \bibinfo {author} {\bibfnamefont {K.}~\bibnamefont {Park}}, \ and\
  \bibinfo {author} {\bibfnamefont {Y.~C.}\ \bibnamefont {Lai}},\ }\href@noop
  {} {\bibfield  {journal} {\bibinfo  {journal} {Phys.\ Rev. E}\ }\textbf
  {\bibinfo {volume} {70}},\ \bibinfo {pages} {035101(R)} (\bibinfo {year}
  {2004})}\BibitemShut {NoStop}%
\bibitem [{\citenamefont {Herrero}(2005)}]{herrero}%
  \BibitemOpen
  \bibfield  {author} {\bibinfo {author} {\bibfnamefont {C.~P.}\ \bibnamefont
  {Herrero}},\ }\href@noop {} {\bibfield  {journal} {\bibinfo  {journal}
  {Phys.\ Rev. E}\ }\textbf {\bibinfo {volume} {71}},\ \bibinfo {pages}
  {016103} (\bibinfo {year} {2005})}\BibitemShut {NoStop}%
\bibitem [{\citenamefont {Zhao}\ \emph {et~al.}(2007)\citenamefont {Zhao},
  \citenamefont {Cupertino}, \citenamefont {Park}, \citenamefont {Lai},\ and\
  \citenamefont {Jin}}]{ZhaoCupertino}%
  \BibitemOpen
  \bibfield  {author} {\bibinfo {author} {\bibfnamefont {L.}~\bibnamefont
  {Zhao}}, \bibinfo {author} {\bibfnamefont {T.~H.}\ \bibnamefont {Cupertino}},
  \bibinfo {author} {\bibfnamefont {K.}~\bibnamefont {Park}}, \bibinfo {author}
  {\bibfnamefont {Y.-C.}\ \bibnamefont {Lai}}, \ and\ \bibinfo {author}
  {\bibfnamefont {X.}~\bibnamefont {Jin}},\ }\href@noop {} {\bibfield
  {journal} {\bibinfo  {journal} {Chaos}\ }\textbf {\bibinfo {volume} {17}},\
  \bibinfo {pages} {043103} (\bibinfo {year} {2007})}\BibitemShut {NoStop}%
\bibitem [{\citenamefont {Huang}\ and\ \citenamefont {Chow}(2009)}]{HuangChow}%
  \BibitemOpen
  \bibfield  {author} {\bibinfo {author} {\bibfnamefont {W.}~\bibnamefont
  {Huang}}\ and\ \bibinfo {author} {\bibfnamefont {T.~W.~S.}\ \bibnamefont
  {Chow}},\ }\href@noop {} {\bibfield  {journal} {\bibinfo  {journal} {Chaos}\
  }\textbf {\bibinfo {volume} {19}},\ \bibinfo {pages} {043124} (\bibinfo
  {year} {2009})}\BibitemShut {NoStop}%
\bibitem [{\citenamefont {Sreenivasan}\ \emph {et~al.}(2007)\citenamefont
  {Sreenivasan}, \citenamefont {Cohen}, \citenamefont {L\'opez}, \citenamefont
  {Toroczkai},\ and\ \citenamefont {Stanley}}]{sreenivasan}%
  \BibitemOpen
  \bibfield  {author} {\bibinfo {author} {\bibfnamefont {S.}~\bibnamefont
  {Sreenivasan}}, \bibinfo {author} {\bibfnamefont {R.}~\bibnamefont {Cohen}},
  \bibinfo {author} {\bibfnamefont {E.}~\bibnamefont {L\'opez}}, \bibinfo
  {author} {\bibfnamefont {Z.}~\bibnamefont {Toroczkai}}, \ and\ \bibinfo
  {author} {\bibfnamefont {H.~E.}\ \bibnamefont {Stanley}},\ }\href@noop {}
  {\bibfield  {journal} {\bibinfo  {journal} {Phys.\ Rev. E}\ }\textbf
  {\bibinfo {volume} {75}},\ \bibinfo {pages} {036105} (\bibinfo {year}
  {2007})}\BibitemShut {NoStop}%
\bibitem [{\citenamefont {Danilla}\ \emph {et~al.}(2006)\citenamefont
  {Danilla}, \citenamefont {Yu}, \citenamefont {Marsh},\ and\ \citenamefont
  {Bassler}}]{danilla2006}%
  \BibitemOpen
  \bibfield  {author} {\bibinfo {author} {\bibfnamefont {B.}~\bibnamefont
  {Danilla}}, \bibinfo {author} {\bibfnamefont {Y.}~\bibnamefont {Yu}},
  \bibinfo {author} {\bibfnamefont {J.~A.}\ \bibnamefont {Marsh}}, \ and\
  \bibinfo {author} {\bibfnamefont {K.~E.}\ \bibnamefont {Bassler}},\
  }\href@noop {} {\bibfield  {journal} {\bibinfo  {journal} {Phys.\ Rev. E}\
  }\textbf {\bibinfo {volume} {74}},\ \bibinfo {pages} {046106} (\bibinfo
  {year} {2006})}\BibitemShut {NoStop}%
\bibitem [{\citenamefont {Danilla}\ \emph {et~al.}(2007)\citenamefont
  {Danilla}, \citenamefont {Yu}, \citenamefont {Marsh},\ and\ \citenamefont
  {Bassler}}]{danilla2007}%
  \BibitemOpen
  \bibfield  {author} {\bibinfo {author} {\bibfnamefont {B.}~\bibnamefont
  {Danilla}}, \bibinfo {author} {\bibfnamefont {Y.}~\bibnamefont {Yu}},
  \bibinfo {author} {\bibfnamefont {J.~A.}\ \bibnamefont {Marsh}}, \ and\
  \bibinfo {author} {\bibfnamefont {K.~E.}\ \bibnamefont {Bassler}},\
  }\href@noop {} {\bibfield  {journal} {\bibinfo  {journal} {Chaos}\ }\textbf
  {\bibinfo {volume} {17}},\ \bibinfo {pages} {026102} (\bibinfo {year}
  {2007})}\BibitemShut {NoStop}%
\bibitem [{Note1()}]{Note1}%
  \BibitemOpen
  \bibinfo {note} {See: http://www.routeviews.org}\BibitemShut {NoStop}%
\bibitem [{\citenamefont {Kleinberg}(2000)}]{kleinberg}%
  \BibitemOpen
  \bibfield  {author} {\bibinfo {author} {\bibfnamefont {J.~M.}\ \bibnamefont
  {Kleinberg}},\ }\href@noop {} {\bibfield  {journal} {\bibinfo  {journal}
  {Nature (London)}\ }\textbf {\bibinfo {volume} {406}},\ \bibinfo {pages}
  {845} (\bibinfo {year} {2000})}\BibitemShut {NoStop}%
\bibitem [{\citenamefont {Arenas}, \citenamefont {Diaz-Guilera},\ and\
  \citenamefont {Guimera}(2001)}]{arenas}%
  \BibitemOpen
  \bibfield  {author} {\bibinfo {author} {\bibfnamefont {A.}~\bibnamefont
  {Arenas}}, \bibinfo {author} {\bibfnamefont {A.}~\bibnamefont
  {Diaz-Guilera}}, \ and\ \bibinfo {author} {\bibfnamefont {R.}~\bibnamefont
  {Guimera}},\ }\href@noop {} {\bibfield  {journal} {\bibinfo  {journal}
  {Physical Review Letters}\ }\textbf {\bibinfo {volume} {86}},\ \bibinfo
  {pages} {3196} (\bibinfo {year} {2001})}\BibitemShut {NoStop}%
\end{thebibliography}

\providecommand{\noopsort}[1]{}\providecommand{\singleletter}[1]{#1}%

\end{document}